\documentstyle[prd,aps,12pt]{revtex}

\input epsf

\begin{document}

\title{Generalized forward scattering amplitudes in QCD at high temperature}

\author{F. T. Brandt and J. Frenkel}
\address{Instituto de F\'\i sica, Universidade de S\~ao Paulo,
S\~ao Paulo, 05389-970 SP, Brazil}

\date{\today}

\maketitle

\vskip 1.0cm

\begin{abstract}
We extend to a general class of covariant gauges an approach which 
relates the thermal Green functions to
forward scattering amplitudes of thermal particles. 
A brief discussion of the non-transversality  of the thermal gluon 
polarization tensor is given in this context. 
This method is then applied to the calculation of the 
${\rm ln }(T)$ contributions associated with general configurations of 
2 and 3-point gluon functions. The results are Lorentz covariant and
have the same structure as the ultraviolet divergent contributions 
which occur at zero temperature.
\end{abstract}


There have been many studies
of the high-temperature behavior of Green functions in thermal QCD
\cite{GrossPisarskiYaffe,Weldon,KajantieKapusta,LandsmanWeert,BraatenPisarski92}.
These investigations have been mainly concerned with hard thermal loops, 
which are important elements for the re-summation of the QCD thermal 
perturbation theory \cite{BraatenPisarski90}. 
The work in reference \cite{Barton}
describes a method for calculating thermal loops, which are expressed
as a momentum integral of
forward scattering amplitudes (summed over spins and
internal quantum numbers) of thermal particles. This approach has been
further elaborated in the Feynman gauge 
and shown to be very useful for determining the partition function
in QCD at high temperature \cite{FrenkelTaylor9092}.

The main purpose of this brief report is to extend the analysis in 
\cite{FrenkelTaylor9092} to a general class of covariant gauges,
characterized by a gauge parameter $\xi$ (the Feynman gauge
corresponds to the special case when $\xi=1$). 
It has been shown that the thermal gluon
polarization tensor is in general not transverse, except in the
Feynman gauge \cite{KajantieKapusta,KobesKunstatterMak,Weldon97}. 
We point out that this property may be simply understood
when one extends, at finite temperature, a procedure which implements
the transversality condition at zero temperature. 
The distinct behavior in the Feynman gauge arises in consequence of
cancellations between ghost and gluon contributions,
which occur in the longitudinal part of the gluon self-energy.
Another application of the above 
approach is the evaluation of the sub-leading ${\rm ln}(T)$
contributions associated with general configurations of 2 and 3-point
gluon functions. After a rather involved 
calculation, we obtain a simple Lorentz covariant result. This
provides a nontrivial verification of a
previous argument \cite{BrandtFrenkel97}, which shows that these
contributions should appear with the same coefficient as the
ultraviolet pole part of the zero temperature amplitudes.

In order to derive the
relation between thermal loops and forward scattering amplitudes
we treat, for definiteness, the case of the gluon self-energy, but
the result generalizes in a obvious way to higher order Green
functions (see Eq. (\ref{eq18})). 
Since the quark loops are independent of the gauge
parameter and have been already considered in \cite{FrenkelTaylor9092},
we shall focus here on the thermal gluon loops.

Consider the contributions associated with the diagram in Fig. 1a, 
where we suppress for simplicity the color indices. We shall employ
an analytic continuation of the imaginary time formalism
\cite{Kapusta}, where these contributions may be written as
\begin{equation}
  \label{eq1}
  \Pi^1_{\mu_1\mu_2}(k,u)=\frac{1}{2\pi i}\int\frac{{\rm d}^3 Q}{\left(2\pi\right)^3}
\int_{\cal{C}}{\rm d Q_0} N\left( Q_0\right)
t^{\alpha\beta\rho\sigma}_{\mu_1\mu_2}\left(Q_0,{\bf Q}, Q_0+k_0,
{\bf Q} + {\bf k}\right)D_{\alpha\beta}(Q) D_{\rho\sigma}(Q+k),
\end{equation}
where $u\equiv(1,0,0,0)$ specifies the rest-frame of the plasma.
The tensor
$t^{\alpha\beta\rho\sigma}_{\mu_1\mu_2}$ is the numerator which is of
degree two in its arguments and $D_{\alpha\beta}(Q)$ is the
bare gluon propagator given by
\begin{equation}
  \label{eq2}
  D_{\alpha\beta}(Q)=\frac{1}{Q^2}\left[\eta_{\alpha\beta}-(1-\xi)
\frac{Q_\alpha Q_\beta}{Q^2}\right] .
\end{equation}
$N(Q_0)=\left[{\rm exp}(Q_0/T)-1\right]^{-1}$ is the Bose
statistical distribution, which has poles along the imaginary 
$Q_0$ axis at $Q_0=2\pi n i T$ ($n=0,\pm 1\, ,\pm 2\,\cdots$). 
The contour $C$ surrounds all the poles of $N(Q_0)$ in a anti-clockwise sense.

  \begin{figure}[htb]
   \hspace{.1\textwidth}

   \vbox{
    \epsfxsize=.8\textwidth
    \epsfbox{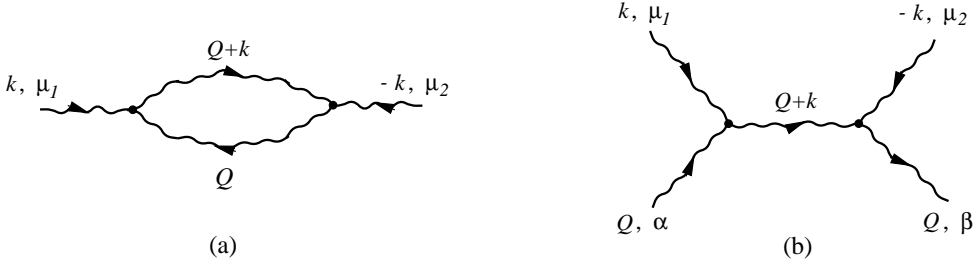}}

   \label{Fig1}
\caption{(a) The gluon self-energy thermal loop. Contributions
from internal ghost particles are to be understood.
(b) An example of the forward scattering graph connected with
diagram (a).}
  \end{figure}

In the contour on the left-hand side of the imaginary axis, we let
$Q_0\rightarrow -Q_0$ and make also the change of variable 
${\bf Q} \rightarrow -{\bf Q}$. Using the relation $N(Q_0)+N(-Q_0)=-1$
and ignoring the temperature independent part, we can write the
thermal contribution of (\ref{eq1}) as
\begin{eqnarray}
  \label{eq3}
\Pi^1_{\mu_1\mu_2}(k,u) & = & 
\frac{1}{2\pi i}\int\frac{{\rm d}^3 Q}{\left(2\pi\right)^3}
\int_{-i\infty+\epsilon}^{i\infty+\epsilon}{\rm d} Q_0 N\left(Q_0\right) 
\nonumber \\
{} & {} & 
\times\left\{
t^{\alpha\beta\rho\sigma}_{\mu_1\mu_2}\left(Q_0,{\bf Q}, Q_0+k_0,
{\bf Q} + {\bf k}\right)D_{\alpha\beta}(Q) D_{\rho\sigma}(Q+k)
+ (Q \rightarrow -Q)
\right\}.
\end{eqnarray}
This can be evaluated with the help of Cauchy's theorem, by encircling
the poles in the Feynman denominators in the right-half plane.
Consider first the poles at $Q_0=|{\bf Q}|$ present in 
$D_{\alpha\beta}\left(Q\right)$.
Evaluating the contributions from simple and double poles, we get
\begin{eqnarray}
  \label{eq4}
\Pi^{1\;a}_{\mu_1\mu_2}(k,u) & = & 
-\int\frac{{\rm d}^3 Q}{\left(2\pi\right)^3}
\left\{\left[\eta_{\alpha\beta}\frac{N(|{\bf Q}|)}{2 |{\bf Q}|}-
\left(1-\xi\right)\frac{\vec{\rm d}}{{\rm d}Q_0}
\frac{Q_\alpha Q_\beta}{\left(Q_0+|{\bf Q}|\right)^2}\right.
N(Q_0)\right]
\nonumber \\
{} & {} & 
\left . \times\left[\frac{ }{ }
t^{\alpha\beta\rho\sigma}_{\mu_1\mu_2}\left(Q_0,{\bf Q}, Q_0+k_0,
{\bf Q} + {\bf k}\right) D_{\rho\sigma}(Q+k)
+ (Q \rightarrow -Q)
\right]\right\}_{Q_0=|{\bf Q}|} ,
\end{eqnarray}
where the derivative $\vec{\rm d}/{\rm d}Q_0$ acts on all terms on its right.
At this point, it is convenient to introduce the operator
\begin{equation}
  \label{eq5}
  P_{\alpha\beta}(Q)\equiv\eta_{\alpha\beta}-\frac{1-\xi}{2}
\frac{\vec{\rm d}}{{\rm d}Q_0}\frac{Q_\alpha Q_\beta}{Q_0}\;\;\; 
\end{equation}
which is related to the sum over the polarization states of the thermal
gluon. In terms of this quantity, equation (\ref{eq4}) may be rewritten as 
\begin{eqnarray}
  \label{eq6}
  \Pi^{1\;a}_{\mu_1\mu_2}(k,u) & = & 
-\frac{1}{\left(2\pi\right)^3}\int\frac{{\rm d}^3 Q}{2|{\bf Q}|}
\left\{P_{\alpha\beta}(Q)N(Q_0) \right .\nonumber \\
{} & {} & 
\times\left . \left[
t^{\alpha\beta\rho\sigma}_{\mu_1\mu_2}\left(Q_0,{\bf Q}, Q_0+k_0,
{\bf Q} + {\bf k}\right) D_{\rho\sigma}(Q+k) + (Q \rightarrow -Q)\right]
\right\}_{Q_0=|{\bf Q}|} .
\end{eqnarray}
Now consider the contributions to $\Pi^{1}_{\mu_1\mu_2}(k,u)$ from
the poles on the right-hand side of the imaginary axis present in 
$D_{\rho\sigma}\left(Q+k\right)$. Since in the imaginary time formalism,
$k_0=2\pi i m T$, such a pole occurs only at $Q_0=-k_0+|{\bf Q}+{\bf k}|$.
Here we make the change
of variable $Q\rightarrow Q-k$ and use the relation
\begin{equation}
  \label{eq7}
  N(Q_0-k_0)=N(Q_0).
\end{equation}
We then get a contribution similar to that in (\ref{eq6}), but with
$k\rightarrow -k$. Thus, the total result for the thermal part of
(\ref{eq1}) is
\begin{eqnarray}
  \label{eq8}
 \Pi^{1}_{\mu_1\mu_2}(k,u) & = & 
-\frac{1}{\left(2\pi\right)^3}\int\frac{{\rm d}^3 Q}{2|{\bf Q}|}
\left\{P_{\alpha\beta}(Q)N(Q_0) \right .\nonumber \\
{} & {} & 
\times\left . \left[
t^{\alpha\beta\rho\sigma}_{\mu_1\mu_2}\left(Q_0,{\bf Q}, Q_0+k_0,
{\bf Q} + {\bf k}\right) D_{\rho\sigma}(Q+k) + (k \rightarrow - k)
+ (Q \rightarrow -Q)\right]
\right\}_{Q_0=|{\bf Q}|} .
\end{eqnarray}
At this stage, {\it after} using (\ref{eq7}), the integrand in
Eq. (\ref{eq8}) is a rational function of $k_0$ which may be
analytically continued to all continuous values of the external
energy.
The first term in the square bracket can be represented by the
forward scattering amplitude 
$A^{\prime\;\alpha\beta}_{\mu_1\mu_2}\left(Q,k,-k\right)$
shown in Fig. 1b. The term containing
$(k \rightarrow - k)$ corresponds to a diagram obtained from 
Fig. 1b by a permutation of the external momenta.

We must also consider the contributions associated with the ghost loop.
Here it is useful to define
\begin{equation}
  \label{eq9}
  \tilde A_{\mu_1\mu_2}^{\alpha\beta}\left(Q,k,-k\right)\equiv
\frac{\eta^{\alpha\beta}}{3+\xi}A_{\mu_1\mu_2}^{ghost}\left(
Q,k,-k\right),
\end{equation}
where $A_{\mu_1\mu_2}^{ghost}\left(Q,k,-k\right)$ represents the
forward-scattering amplitude of a ghost particle by the external
gluon fields. 
The above definition is convenient
because of the identity $P_{\alpha\beta}\tilde A_{\mu_1\mu_2}^{\alpha\beta}
=A_{\mu_1\mu_2}^{ghost}$, which holds due to the fact that the ghost particle
is on-shell. 

In order to obtain the full thermal amplitude for the gluon self-energy,
we must include the contributions from the tadpole graph shown in Fig. 2.
Denoting by $A_{\mu_1\mu_2}^{\alpha\beta}(Q,k,-k)$ the total forward
scattering amplitude, we can express the complete thermal contribution
in terms of a momentum integral of $A_{\mu_1\mu_2}^{\alpha\beta}$, as
\begin{equation}
  \label{eq10}
 \Pi_{\mu_1\mu_2}(k,u) =  
-\frac{1}{\left(2\pi\right)^3}\int\frac{{\rm d}^3 Q}{2|{\bf Q}|}
\left\{P_{\alpha\beta}(Q)N(Q_0)
\left[
A^{\alpha\beta}_{\mu_1\mu_2}\left(Q,k,-k\right)+ (Q \rightarrow -Q)
\right]\right\}_{Q_0=|{\bf Q}|} .
\end{equation}
Using this expression, it can be verified that for the exact gluon
self-energy, $k^{\mu_1} \Pi_{\mu_1\mu_2}\neq 0$, a property that was
previously investigated \cite{KajantieKapusta,KobesKunstatterMak,Weldon97}.
Multiplying the forward scattering amplitude in Eq. (\ref{eq10}) by
$k^{\mu_1}$, we find that the longitudinal part of the polarization
tensor can be expressed as
\begin{eqnarray}
  \label{eq13}
k^{\mu_1}\Pi_{\mu_1\mu_2} & = &\left(1-\xi\right)  
\frac{g^2C_G}{\left(2\pi\right)^3}\int\frac{{\rm d}^3 Q}{2|{\bf Q}|}
\left\{
\left(
\frac{k^2}{k^2+2k\cdot Q}+\frac 1 2 
\frac{\vec {\rm d}}{{\rm d}Q_0} \frac{k\cdot Q}{Q_0}\right) \right. 
\nonumber \\
{} & {} & \left.\;\;\;\;\;\;\;\;\;\;\;\;\;\;\;\;\;\;\;\;\;\;
\;\;\;\;\;\;\;\;\;\;\times
\;\;\frac{k\cdot Q k_{\mu_2}-k^2 Q_{\mu_2}}{k^2+2k\cdot Q}N(|Q_0|)
 + (Q \rightarrow -Q)
\right\}_{Q_0=|{\bf Q}|} ,
\end{eqnarray}
where $C_G$ is the gluon Casimir invariant.
This relation shows that the thermal part of 
$\Pi_{\mu_1\mu_2}$ is transverse only in the Feynman gauge (but
$k^{\mu_1} k^{\mu_2} \Pi_{\mu_1\mu_2}=0$ for all values of $\xi$).

  \begin{figure}[htb]
   \hspace{.1\textwidth}

   \vbox{
    \epsfxsize=.65\textwidth
    \epsfbox{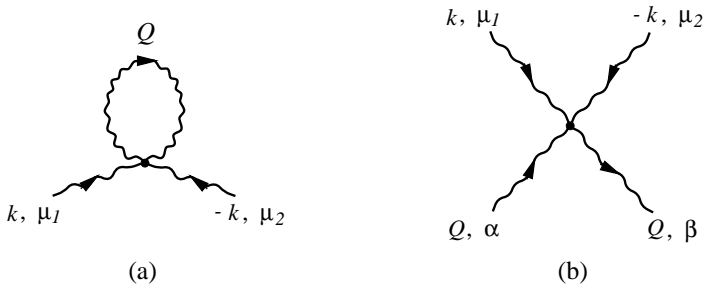}}

   \label{Fig2}
\caption{(a) The tadpole loop diagram and the corresponding 
forward-scattering graph(b).}
  \end{figure}

There is an interesting connection between this behavior and the
transversality condition which must hold at zero temperature. To see
this, we recall that the integrand appearing in the self-energy function
at finite temperature is similar, apart from the Bose factor, to the
one which occurs at zero temperature \cite{Kapusta}. 
Thus, we are lead to consider
the way that the zero temperature property $k^{\mu_1}\Pi_{\mu_1\mu_2}=0$
is actually implemented at the one-loop order. In the Feynman gauge, due to 
cancellations between gluon and ghost loops, this condition is verified
just by making a shift $Q\rightarrow Q-k$ in some terms in the 
integrand. Such a shift can be performed at zero temperature using 
the dimensional regularization scheme 
and is also allowed at finite temperature in the imaginary time
formalism, in view of property (\ref{eq7}). 
In a general covariant gauge, the ghost
contribution remains the same since it is independent of the gauge
parameter $\xi$. But the longitudinal part of the gluon loop contribution
differs at the integrand level from that obtained in the Feynman gauge
by terms proportional to $(1-\xi)$. In this case, the verification
of the transversality condition at zero temperature requires, in addition, 
the use of dimensionally regularized integrals like
\begin{equation}
  \label{eq14}
\int\frac{{\rm d}^n Q}{\left(Q^2\right)^\alpha} f\left(Q\right)=0 .
\end{equation}
However, such a result does not hold at finite temperature when
$f(Q)=N({Q_0})$, in which case (\ref{eq14}) becomes instead proportional
to $T^{n-2\alpha}$. Consequently, the above longitudinal part of
the gluon loop contribution will not vanish at finite temperature,
in agreement with the result given by equation (\ref{eq13}).

Let us next consider the contributions of (\ref{eq10}) coming from
the hard thermal region where $|{\bf Q}|\gg |{\bf k}|,k_0$. 
This region is relevant for the determination
of $T^2$ and ${\rm ln}(T)$ contributions (unlike the terms linear 
in $T$ which come also from soft momenta).
We will
argue that in this case we can effectively commute the differential
operator $P_{\alpha\beta}$ and the Bose factor $N(Q_0)$. To
this end, let us study the contribution obtained when the derivative
from $P_{\alpha\beta}$ acts on $N(Q_0)$. This leads to an integrand
proportional  to a factor $N^{\prime}(|{\bf Q}|)/|{\bf Q}|^2$ times
$Q_\alpha Q_\beta A^{\alpha\beta}_{\mu_1\mu_2}$, where the thermal
particle four-vector $Q$ is on shell. As a consequence of the gauge invariance
of the forward scattering amplitude, 
$Q_\alpha Q_\beta A^{\alpha\beta}_{\mu_1\mu_2}$ vanishes when the external
gauge fields are on shell and transverse. It is not difficult to show
that even for general values of $k$ there will be strong cancellations,
so that 
$Q_\alpha Q_\beta A^{\alpha\beta}_{\mu_1\mu_2}$
actually becomes a function of zero degree in $Q$. Together with the
above factor, such a function will produce in (\ref{eq10}) an integral
of the form
\begin{equation}
  \label{eq13n}
\int_\tau^\infty{\rm d}|{\bf Q}|N^\prime(|{\bf Q}|)=
\frac{1}{1-{\rm exp}(\tau/T)},
\end{equation}
where the lower limit $\tau\gg|{\bf k}|,k_0$ delimitates the hard
thermal region. 
Since (\ref{eq13n}) does not yield at high temperature
$T^2$ or ${\rm ln}(T)$ contributions, it may be neglected
for our purpose so that (\ref{eq10}) becomes equivalent to
\begin{equation}
  \label{eq12}
 \Pi_{\mu_1\mu_2}^{{\rm ht}}(k,u) =  
-\frac{1}{\left(2\pi\right)^3}\int\frac{{\rm d}^3 Q}{2|{\bf Q}|}
N(|{\bf Q}|)\left\{P_{\alpha\beta}(Q)
A^{\alpha\beta}_{\mu_1\mu_2}\left(Q,k,-k\right)+ (Q \rightarrow -Q)
\right\}_{Q_0=|{\bf Q}|} .
\end{equation}
We may now extract from (\ref{eq12}) a series of high temperature terms
which come from the region of large $Q$.
Since the four momentum
of the thermal particle is ultimately set on shell, the denominators
in $A_{\mu_1\mu_2}^{\alpha\beta}$ can be expanded as follows
\begin{equation}
  \label{eq15}
\left.\frac{1}{\left(k+Q\right)^2}\right|_{Q_0=|{\bf Q}|}
= \frac{1}{2 k\cdot Q}-\frac{k^2}
{\left(2 k\cdot Q\right)^2}+\cdots .
\end{equation}
We may also expand the numerator in powers of $k_\mu/|{\bf Q}|$.
The first term has a denominator  of the form $(k\cdot Q)^{-1}$
and a numerator quadratic in $Q$ which is independent of $k$. But such terms
cancel out in (\ref{eq12}) by symmetry under $Q\rightarrow -Q$.
(For the same reason, any terms odd in Q will generally cancel out).
The next contributions are down by a power of $k_\mu/|{\bf Q}|$ and
arise from those terms in $A_{\mu_1\mu_2}^{\alpha\beta}$ which are
of zero degree in $Q$. Such terms give the well known leading $T^2$
contributions, which are gauge independent. The following non-vanishing
contributions arise from those terms in $A_{\mu_1\mu_2}^{\alpha\beta}$
which are of degree $-2$ in $Q$. By power counting, such terms
will produce ${\rm ln}(T)$ contributions. 
The corresponding term in (\ref{eq12}) coming from the trace
$A^{\alpha\alpha}_{\mu_1\mu_2}$ yields, after performing the
Q-integration, a Lorentz covariant expression. The other
part of the integrand involves the operator 
$\vec{\rm d}/{\rm d}Q_0 \, Q_0^{-1}$ applied to a covariant function of
zero degree in $Q$. Such a function can be generated by 
differentiating ${\rm ln}(k\cdot Q)$ with respect to $k_{\mu_i}$.
We are thus lead to consider the integral
\begin{equation}
  \label{eq16n}
\int \frac{{\rm d}^3 Q}{|{\bf Q}|} N(|{\bf Q}|)\left\{
\frac{{\rm d}}{{\rm d} Q_0}\left(\frac{1}{Q_0} {\rm ln}(k\cdot Q)\right)
\right\}_{Q_0=|{\bf Q}|}=
\pi {\rm ln}(T) {\rm ln}(k^2)+\cdots ,
\end{equation}
where we have written explicitly only the $k$-dependent
${\rm ln}(T)$ term. Then the ${\rm ln}(k^2)$ term will generate, 
after differentiations with respect to $k_{\mu_i}$, a Lorentz
covariant expression. Adding all such contributions, we obtain
for the logarithmic dependence on the temperature the simple result
\begin{equation}
  \label{eq16}
\Pi_{\mu_1\mu_2}^{{\rm ln}}=-\left(\frac{13}{3} - \xi \right)
\frac{g^2 C_G}{\left(4\pi\right)^2} {\rm ln}\left(T\right)
\left(\eta_{\mu_1\mu_2} k^2 - k_{\mu_1}  k_{\mu_2}\right) . 
\end{equation}
The important point about (\ref{eq16}) is
the fact that it has the same structure as the ultraviolet
divergent contribution of the gluon self-energy at zero temperature
\cite{Muta}
\begin{equation}
  \label{eq17}
\Pi_{\mu_1\mu_2}^{{\rm UV}}=\left(\frac{13}{3} - \xi \right)
\frac{g^2 C_G}{\left(4\pi\right)^2}
\left(\frac{1}{2\epsilon}+{\rm ln}\left(M\right)\right)
\left(\eta_{\mu_1\mu_2} k^2 - k_{\mu_1}  k_{\mu_2}\right) , 
\end{equation}
where the space-time dimension is $4-2\epsilon$ and $M$ is the
renormalization scale.

The result (\ref{eq10}) for the thermal gluon self-energy can be generalized
in a straightforward way to higher order n-point gluon functions. 
In this case, let us denote by
$A_{\mu_1\mu_2\cdots\mu_n}^{\alpha\beta}\left(Q,k_1,k_2,\cdots,k_n\right)$
the forward scattering amplitude of a thermal particle with
momentum $Q$ by external fields with momenta $k_1,k_2,\cdots,k_n$
such that $k_1+k_2+\cdots+k_n=0$.  
(We have omitted, for simplicity of notation, the color indices).
The thermal part of the
one-loop n-point gluon function can then be expressed in terms of a
momentum integral of the forward scattering amplitude as
\begin{equation}
  \label{eq18}
 \Gamma_{\mu_1\cdots\mu_n}(k_1,\cdots,k_n,u) =  
-\frac{1}{\left(2\pi\right)^3}\int\frac{{\rm d}^3 Q}{2|{\bf Q}|}
\left\{P_{\alpha\beta}(Q)N\left(Q_0\right)\left[
A^{\alpha\beta}_{\mu_1\cdots\mu_n}
\left(Q,k_1,\cdots,k_n\right)+ (Q \rightarrow -Q)
\right]\right\}_{Q_0=|{\bf Q}|} ,
\end{equation}
where $u$ is the four-velocity of the heat bath, 
$P_{\alpha\beta}(Q)$ indicates the operator defined in (\ref{eq5})
and $N\left(Q_0\right)$ is the Bose distribution function.

Let us apply this result to the evaluation of the hard thermal contributions
associated with the 3-point gluon function. The corresponding forward
scattering amplitude is represented in Fig. 3, where graphs
obtained by all permutations of external momenta and indices are to be
understood. Proceeding in a similar
way to that indicated following Eq. (\ref{eq12}), 
one finds that the terms of zero degree in $Q$ from
$A^{\alpha\beta}_{\mu_1\mu_2\mu_3}(Q,k_1,k_2,k_3)$ yield the well
known leading $T^2$ contributions, which are gauge invariant.
The terms from 
$A^{\alpha\beta}_{\mu_1\mu_2\mu_3}(Q,k_1,k_2,k_3)$
which are of degree -2 in $Q$ give the high temperature
${\rm ln}(T)$ contributions. After a very involved computation
which uses an extension of the approach described in 
\cite{BrandtFrenkel97}, we obtain for these contributions the
simple Lorentz covariant result
\begin{equation}
  \label{eq19}
\Gamma_{\mu_1\mu_2\mu_3}^{{\rm ln}}=
- \left(\frac{17}{6} - \frac{3\xi}{2} \right)
\frac{g^2 C_G}{\left(4\pi\right)^2} {\rm ln}\left(T\right)
i\left(\eta_{\mu_1\mu_2}\left(k_1-k_2\right)_{\mu_3}+
       \eta_{\mu_2\mu_3}\left(k_2-k_3\right)_{\mu_1}+
       \eta_{\mu_3\mu_1}\left(k_3-k_1\right)_{\mu_2} 
\right) .
\end{equation}
This is proportional to the basic 3-gluon vertex and has the same
structure as the ultraviolet divergent part of the 3-point gluon
function at zero temperature, which is given \hbox{by \cite{Muta}}
\begin{equation}
  \label{eq20}
\Gamma_{\mu_1\mu_2\mu_3}^{{\rm UV}}=
 \left(\frac{17}{6} - \frac{3\xi}{2} \right)
\frac{g^2 C_G}{\left(4\pi\right)^2} 
\left(\frac{1}{2\epsilon}+{\rm ln}\left(M\right)\right)
i\left(\eta_{\mu_1\mu_2}\left(k_1-k_2\right)_{\mu_3}+
       \eta_{\mu_2\mu_3}\left(k_2-k_3\right)_{\mu_1}+
       \eta_{\mu_3\mu_1}\left(k_3-k_1\right)_{\mu_2} 
\right) . 
\end{equation}

  \begin{figure}[htb]
   \hspace{.1\textwidth}

   \vbox{
    \epsfxsize=.8\textwidth
    \epsfbox{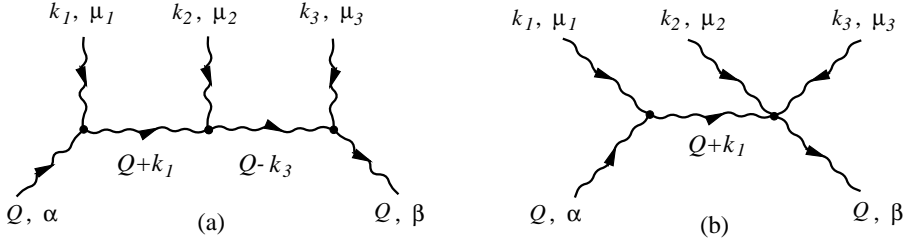}}

   \label{Fig3}
\caption{Examples of forward scattering amplitudes associated with
the thermal 3-gluon loop diagrams. Graphs involving the forward
scattering of ghost particles should be included.}
  \end{figure}

We note that the ${\rm ln}(T)$ in Eqs. (\ref{eq16}) and
(\ref{eq19}) may be combined with the ${\rm ln}(M)$ from
Eqs. (\ref{eq17}) and (\ref{eq20}) respectively, to yield ${\rm ln}(T/M)$ 
contributions. Although these terms have gauge-dependent coefficients,
the quantity
\begin{eqnarray}
  \label{eq21}
Z^{-1}_{g}\left(T/M\right) & \equiv &
\left[1-\left(\frac{13}{3}-\xi\right)\frac{g^2C_G}{\left(4\pi\right)^2}
{\rm ln}\left(\frac T M\right)\right]^{3/2}
\left[1-\left(\frac{17}{6}-\frac{3\xi}{2}\right)
\frac{g^2C_G}{\left(4\pi\right)^2}
{\rm ln}\left(\frac T M\right)\right]^{-1} \nonumber \\
{} & \simeq &
\left[1+\frac{11 g^2 C_G}{48 \pi^2} {\rm ln}\left(\frac T M\right)
\right]^{-1}
\end{eqnarray}
is gauge invariant. This factor, which sums up the 
one-loop
${\rm ln}\left(T \right)$ contributions to the running coupling constant
$\bar g(T)$ at high temperature, 
is relevant, for example, in the calculation
of the pressure in thermal QCD \cite{ArnoldZhai,BraatenNieto}.

\acknowledgements{We would like to thank CNPq (Brasil) for a grant. J. F. is 
grateful to Prof. J. C. Taylor for helpful discussions.}


\end{document}